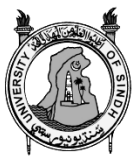
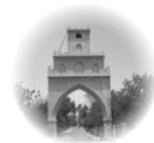

# SINDH UNIVERSITY RESEARCH JOURNAL (SCIENCE SERIES)

## Secure Network in Business-to-Business application by using Access Control List (ACL) and Service Level Agreement (SLA)

A. A. ABRO, S. SOOMRO++, Z. ALANSARI, M. R. BELGAUM, B. K. KHAKWANI*

College of Computer Studies, AMA International University, Bahrain.

**Abstract:** The motivation behind this paper is to dissecting the secure network for Business to Business (B2B) application by Implementing Access Control List (ACL) and Service Level Agreement (SLA). This data provides the nature of attacks reported as external or internal attacks. This paper presents the initial finding of attacks, types of attacks and their ratio within specific time. It demonstrates the advance technique and methodology to reduce the attacks and vulnerabilities and minimize the ratio of attacks to the networks and application and keep the network secure and runs application smoothly regarding that. It also identifies the location of attacks, the reason behind the attack and the technique used in attacking. The whole field of system security is limitless and in an evolutionary stage. To comprehend the exploration being performed today, foundation learning of the web and assaults, the security is vital and in this way they are investigated. It provides the statistical analytics about various attacks and nature of attacks for acquiring the results through simulation to prove the hypothesis.

## 1. INTRODUCTION

System security begins with confirming, for example, username and a secret word. The world is getting to be more interconnected with the appearance of the web and new security advances presented (Taubenberger, *et al*., 2012). There is a huge number of individual, business, military; government data is accessible over the web around the world. Security is getting to be more criticalness in light of licensed innovation that could be effortlessly procured through over the web (Kesh, *et al*., 2002). Once verified, access policy firewall authorizes, e.g. what access permitted to the client's systems. It is prevailing to counteract of unpermitted access, and the data may omit to check conceivably unsafe stuff. As an example, infected systems during communication transmit worms, Trojans over other systems. The intrusion prevention system helps to identify and repress the malware activities of hostile infected programming. The irregular congestion or interruption location framework may screen the wire shark traffic that might log for review or examine for an abnormal state. Two hosts or clients utilizing system may keep up privacy by correspondence. Many formal principles and other specialized details and programming characterize the operation of different parts of the www, the Web, and machine data trade. Security plays important key role in whole scenario, security and its parameter for the network and group policy object impose better security through that.

In the Digital Era, the security is a very significant factor and very essential for all organization (Niranjanamurthy, 2013). The parameter of security is designed according to the obstacles and challenges. Security parameter is different for different type of scenario (Li, *et al*., 2011). Security is different for all kinds of situations. Access Control list with the support of Service level agreement performs the better security and only authorize user can access and enter the network and access the information.

Table-1: Attack method and Security Technology

| PC Security Characteristics | Assault Method | Innovation for web Security |
|---|---|---|
| **Discretion** | Spying, Hacking, Phishing, Dos and IP Spoofing | IDS, Firewall, Cryptographic, Framework, IPSec and SSL |
| **Reliability** | Worms, Trojans, Eavesdropping, DoS and IP Spoofing | IDS, Firewall, Anti-Malware, Software, IPSec and SSL |
| **Secrecy** | Email, Spamming, Hacking, DoS and Cookies | IDS, Firewall, Anti-Malware, Software, IPSec and SSL |
| **Convenience** | DoS, Email, Bombing, Spamming and Systems Boot Record Infectors | IDS, Firewall, Anti-Malware, Software, IPSec and SSL |

In this paper, we talk about the core standards for customer perception and its execution lists, and the

++Corresponding author: Safeeullah Soomro   email: s.soomro@amaiu.edu.bh
* Department of General Studies, Jubail Industrial College KSA

outline and usage of the implementation assessment framework (Marilly, 2002). Through this way, the confidentiality and security keep and due to the implementation of service level agreement the third party directly involve to provide the security at their end side to maintain the secure environment (Ausanka-Crues, 2001). Furthermore, the parameter is frequently passed from one portal to another portal and user is facilitate often and safe portal provides the dynamic integration of product as well. The home based system or small office based system immediately required the security essentials on other hands generally large organizations may need high support and advancement of programming and latest equipment to keep malevolent assaults from hacking and spamming.

## 2. BACKGROUND

Network Security is the possibilities for network security are nearly endless since it comprises of procurement and arrangements which actualize by system controlling authority avoid and through screen unapproved access, change, system open assets, and abuse. The system security needed the approval of access for information by system overseer. System security blankets both open and private machine system (Kaushik, 2012). The attacks such as DDos attack, TCP hijacking, the vulnerability like eavesdropping, exploits like Trojans, malware, virus and payloads like root kits and key loggers, using software bugs, buffer overflows, packet sniffing and intrusion detection systems and their countermeasures (Alexliu, *et al*., 2011). While concerning privacy the email hacking; spamming; bombing; DoS; and Cookies are the attacks, but their internet securities are Firewall; Anti-Malware; IDS; IPS; IPSec; and SSL (Abadi, *et al*., 2011). Whereas in accessibility are concerned assaults are Email bombarding, DoS; Spamming and Boot framework and Record Infectors and their engineering of web security are IDS, Anti-Malware, Software, and Firewall. Honesty concern are infections, DoS; worms; Trojans; IP Spoofing; and Eavesdropping, on another hand its web securities are Firewall; Anti-Malware; IDS; IPSec; and SSL. Privacy concerned assaults are DoS; IP Spoofing; Phishing; Eavesdropping; Hacking, and their relevant securities are Firewall; Cryptographic Systems; IDS; IPSec; and SSL.

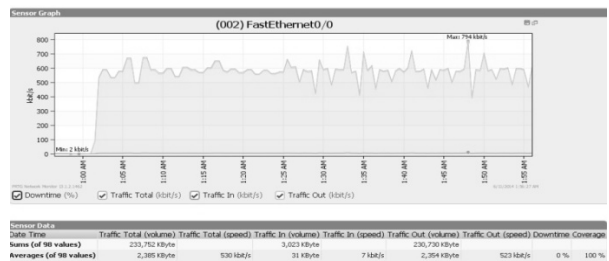

**Fig-2 Network Security with different security Function**

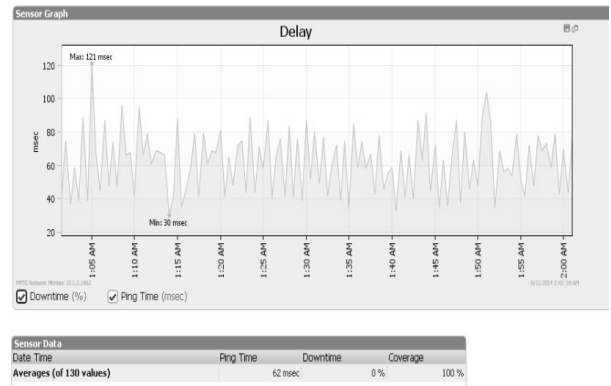

**Fig-3 Delay of Physical Network**

## 3. MATERIALS AND METHODS

Following are the performance graphs captured on the monitoring server as a result of sending packets from the source to the respective destinations. The performance and evaluation of the technique are measured through convergence time, physical network and security function. This methodology performance based on secure routing of the user and control its activity beyond of that most sophisticated flow of information and transaction through a different route and making it more secure way of transaction. Many different attacks are being encountered due to the movement of the operation. In this methodology, we measure framework ease of use and execution through systems introduced to the customers. The reason for our framework is to clarify the conduct of the client applications that permits us to quantify framework execution for customer ease of use. The tool can able the throughput and reaction speed toward the end-point applications, which affects the client. The execution assessment evaluations expect to enhance the execution of the endpoint application. The aggregate framework execution doesn't generally interrelate with the execution of the system way. In this way, the framework system execution ought to be assessed in the customer applications.

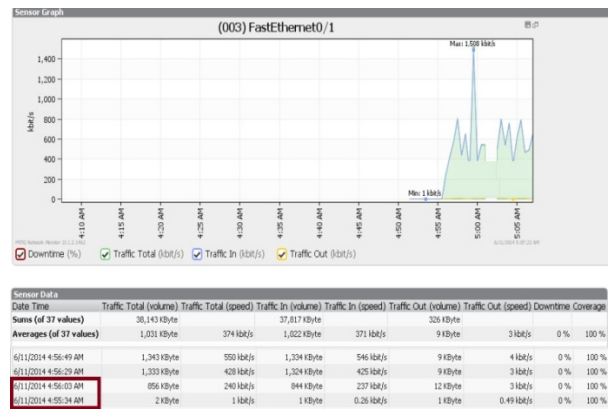

**Fig-4 Convergence time**

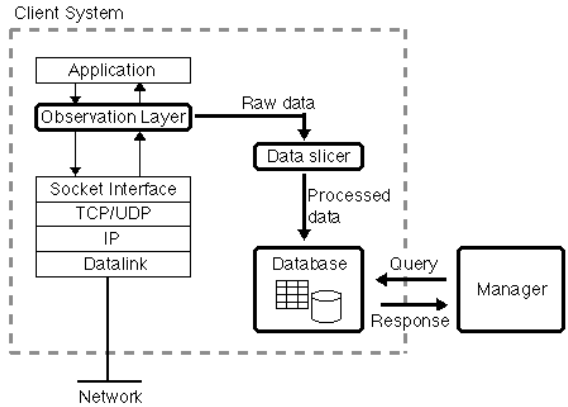

**Fig-5 Model of Performance Evaluation**

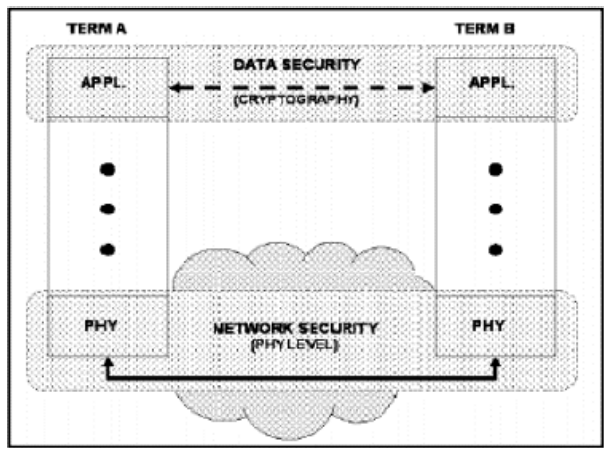

**Fig-6 Model of Performance Evaluation**

### 4. DISCUSSION

The framework of system security depends on the layers of insurance which comprise system administration and safety system programming, but it does not depend on upon the system devices or equipment. To ensure the relevant result as far as Network Security, the unapproved clients get to the assets and rupture the security. The privacy and security are extremely vital to guarantee the system security. System security is extremely critical field as its undeniably spontaneous consideration of web and stretches. The security innovations, for the most part, software based however various fittings are likewise utilized for that purpose. Network security is vast field as it is considered as the web grows. The security innovation focused on security hazard and web conventions. Security innovation is ordinarily product yet fittings are additionally included in that.

with secure access. Some attacks and as well as critical vulnerabilities can be reduced, and performance with user satisfaction can be achieved through this network. This method follows the advance security standards, which are mostly used by corporates where a high level of business can be moved through transactions process. In this methodology there are many rich source of data can be utilized with advanced method many attacks are denied at primary defence which are not countable sometime.

Moreover, this technique is very easy to manage the user information and control the user accessibility. Well, the web security more often than not actualizes the new web convention Ipv6 may give numerous web profits to clients. The system security may need to develop all the more quickly manage the dangers and vulnerabilities further later on.

### 5. CONCLUSION

The security is the main area which is having space or flaws for the improvement and enhancement. Regarding this, the researcher would like to share ideas through several research projects we can improve new techniques of security. In future, the actual and different parameter of security will be introduced and as well as latest technologies regarding third parties' security providers also finishing and adding more advanced and latest way of a secure network from an unauthorized user and their illegal activity in network defiantly not supposed to be in favour of the organization. Security in the digital era is becoming very essential and key factor through which only authorize user can access the resources, data, and information. The Immune system will fight off attacks and allows only the trusted users. The future will possibly be that the security is similar to an immune system. Improvement and enhancement is a continuous process which will lead to the perfection and betterment of safety, confidentiality and security proving better for the network and application.